\documentclass[prl,twocolumn,amsmath,amssymb,showpacs,superscriptaddress]{revtex4-1}

\usepackage{amsfonts}
\usepackage{amssymb}
\usepackage{amsmath}
\usepackage{graphicx}
\usepackage{soul,xcolor}
\usepackage{dcolumn}
\usepackage{bm}
\usepackage[T1]{fontenc}
\usepackage{hyperref}

\begin{document}

\title{Cross correlations in disordered, four-terminal graphene-ribbon conductor: Hanbury-Brown and Twiss
exchange as a sign of non-universality of noise }

\author{Z. Tan}
\affiliation{Low Temperature Laboratory, Department of Applied Physics, Aalto University, Espoo, Finland}
\author{T. Nieminen}
\affiliation{Low Temperature Laboratory, Department of Applied Physics, Aalto University, Espoo, Finland}
\author{A. Puska}
\affiliation{Low Temperature Laboratory, Department of Applied Physics, Aalto University, Espoo, Finland}
\author{J. Sarkar}
\affiliation{Low Temperature Laboratory, Department of Applied Physics, Aalto University, Espoo, Finland}
\author{P. L\"ahteenm\"aki}
\affiliation{Low Temperature Laboratory, Department of Applied Physics, Aalto University, Espoo, Finland}
\author{F. Duerr}
\affiliation{Physikalisches Institut (EP3), University of W\"urzburg, W\"urzburg, Germany}
\author{C. Gould}
\affiliation{Physikalisches Institut (EP3), University of W\"urzburg, W\"urzburg, Germany}
\author{L. W. Molenkamp}
\affiliation{Physikalisches Institut (EP3), University of W\"urzburg, W\"urzburg, Germany}
\author{K. E. Nagaev}
\affiliation{Kotelnikov Institute of Radioengineering and Electronics, Russian Academy of Science,  Moscow, Russia}
\affiliation{Moscow Institute of Physics and Technology, Dolgoprudny, Russia}
\author{P. J. Hakonen}\email[Corresponding author: pertti.hakonen@aalto.fi]{}
\affiliation{Low Temperature Laboratory, Department of Applied Physics, Aalto University, Espoo, Finland}

\begin{abstract}

We have investigated current-current correlations in a cross-shaped  conductor made of graphene ribbons.  We  measured auto and cross correlations and compared them with the theoretical predictions for ideal diffusive conductors. Our data deviate from these predictions and agreement can be obtained only by adding contributions from occupation-number noise in the central region connecting the arms of the cross. Furthermore, we have determined Hanbury -- Brown and Twiss (HBT) exchange correlations in this system. Contrary to expectations for a cross-shaped diffusive system, we find finite HBT exchange effects due to the occupation-number noise at the crossing. The strength of these HBT exchange correlations is found to vary with gate voltage, and very a distinct  HBT effect with large fluctuations is observed near the Dirac point.

\end{abstract}


\maketitle


Disordered graphene is an extraordinary tunable system for studying electrical conduction ranging from nearly ballistic transport \cite{miao2007,danneau2008} to hopping conductivity  \cite{chen2007,han2007,ozyilmaz2007,Han2010,oostinga2010, danneau2010}. In narrow graphene wires, in particular, the number of transport channels can be varied significantly by tuning charge density by gate voltage and conduction can be pinched off fully  near the charge neutrality point (CNP). The elastic mean free path can be maintained relatively large compared with device dimensions, while the importance of localization and Coulomb interactions can be varied by adjusting the charge density \cite{sols2007,Droscher2011,ensslin2012}. Disorder in graphene can lead either to increase or decrease of shot noise \cite{prada2007,lewenkopf2008,mirlin2010,mucciolo2010}. Thus, in graphene nanoribbon (GNR) systems, it is possible to study physics of current-current correlations in a regime where disorder can be tuned, which makes it an excellent platform for investigating  noise properties of disordered conductors.

Shot noise originates from the granular nature of charge carriers, and it can be used as an independent test for the conduction mechanism \cite{Blanter2000,lesovik2011}. However it is difficult to distinguish
between different models of noise in graphene
using two-terminal measurements because several of them give nearby strength, on the order
of 0.3 - 0.4, when compared to Poissonian noise. One of the ways to overcome this difficulty is measuring the cross-correlated noise in multiterminal
graphene systems.

Theoretical treatment of current-current cross correlations $S_{nm}$  between terminals $n$ and $m$  in a diffusive cross geometry as well as  their relation to auto correlations has been performed in Refs. \onlinecite{buttiker1997} and \onlinecite{Sukhorukov1999} with virtually equivalent findings. In the semiclassical theory  \cite{Sukhorukov1999}, the spectral density of noise in a diffusive system is governed by  the local distribution function. This function is sensitive to diffusion of electrons, which is dependent on the local conductance and geometry of the conductor.
The semiclassical theory predicts similar  behavior for shot noise in all cross-shaped diffusive conductors with negligible resistance of the central region. The Fano factor, \textit{i.e.}  the ratio between autocorrelation noise power $S_{nn}$ and the Poissonian noise $S_P=eI_n$ related to current $I_n$ in terminal n,  is found to remain at $F=1/3$, \textit{i.e.} as for a single wire, when biasing is done at terminal 1 and other terminals are grounded. In particular, the semiclassical theory predicts additivity of cross correlations in such a cross-shaped conductor, which would mean the absence of Hanbury--Brown and Twiss (HBT) exchange effects \cite{buttiker1997} in our sample.

In this Letter, we  report measurements on auto and cross correlations which are compared  with the theoretical
predictions for ideal diffusive conductors \cite{buttiker1997,Sukhorukov1999,Blanter2000}. Our data deviate from the
 noise predictions for diffusive systems and agreement can be obtained only by adding contributions
from occupation-number noise in the central region connecting the arms of the cross. The presence of this
noise is in line with observed negative bend resistance, which indicates partly ballistic transmission across the
centre of our graphene sample. Our experiments demonstrate the existence of HBT exchange effects with clear
difference from the expectations for a regular diffusive system.  The HBT effect varies substantially with gate
voltage and it becomes very strong near the CNP.

In a graphene cross device, there may be deviations out from the framework of the semiclassical theory. The basic
assumption of diffusive transport theory is that the mean free path $\ell_{mfp} \ll \min\{L,W\}$ compared with the
dimensions of the sample, the latter condition of which is not well fulfilled in a narrow GNR. Deviations from finite
size effects are estimated in Ref. \onlinecite{buttiker1997}, which predicts a small positive exchange term on the
order of $(\ell_{mfp}/L)\,G_0\,eV$ for a metallic diffusive cross. This prediction,  however, has the opposite
sign with respect to our experimental results, which are more reminiscent of the behavior of a multiterminal chaotic
quantum dot \cite{VanLangen1997}. Our measurements, in fact, reveal non-local conductance which indicates that
$\ell_{mfp} \simeq W$ and that the transport over the central area of the cross for some charge carriers is
ballistic. Furthermore, strong disorder scattering may change the Fano-factor in graphene. Our results show that the
noise properties of the system can be accounted for by the standard diffusive theory
provided that the central region is considered as a four-terminal ballistic conductor with nonequilibrium electron
distributions in the terminals, which also contributes to the noise through the occupation-number fluctuations of incident electrons \cite{landauer1992}.
\begin{figure}[tbp]
\includegraphics[width=5.5cm]{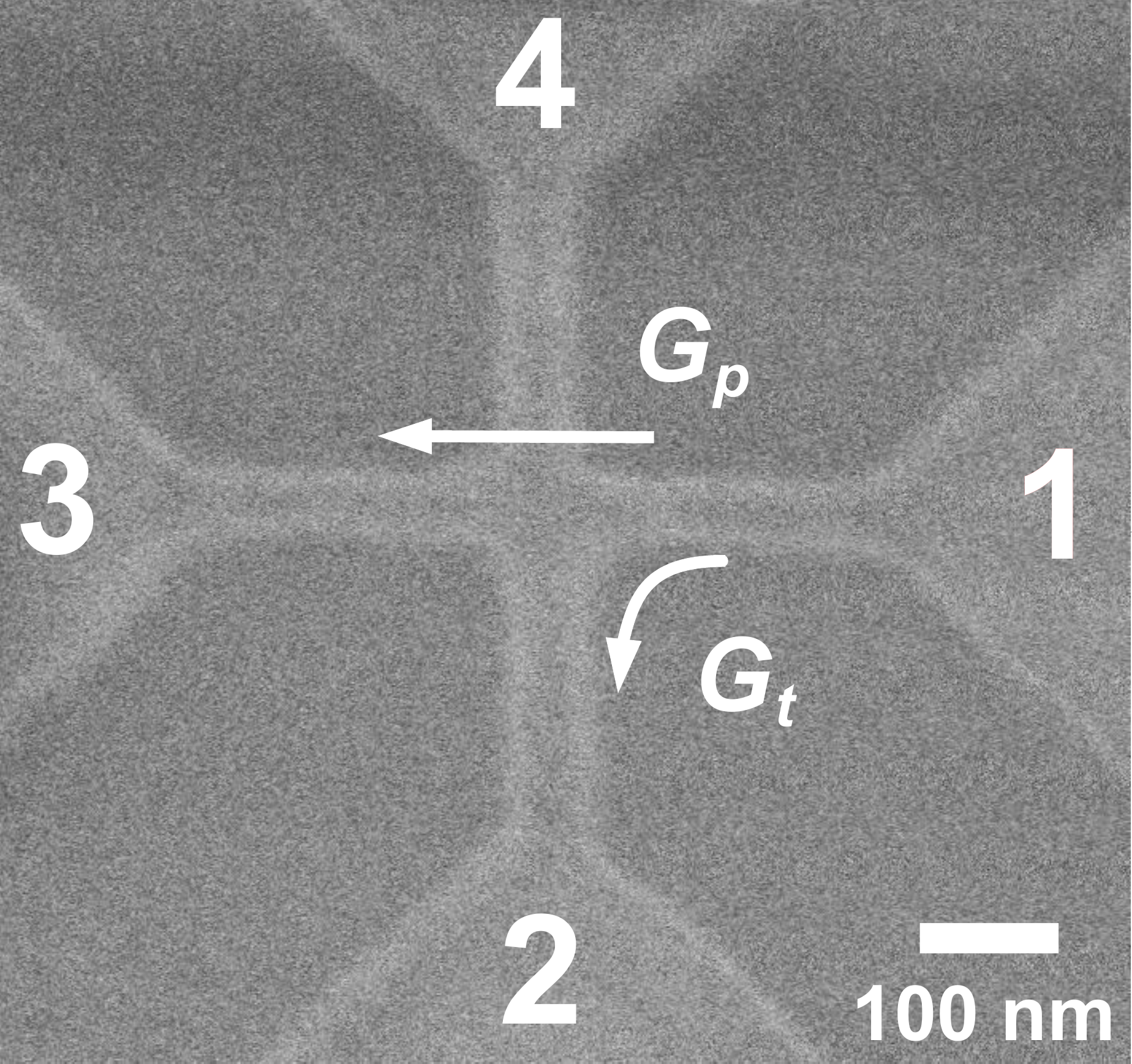}
\caption{ Scanning electron micrograph of the GNR sample; light gray denotes the edges of the graphene. Terminals 1 and 3 were employed for cross correlation while bias was supplied via 2 and 4 in the HBT experiments. The white scale bar corresponds to 100 nm. The overlaid arrows define the straight and bent carrier paths with conductances of $G_p$ and $G_t$, respectively. }
\label{sample}
\end{figure}

 We investigated current-current cross correlations in a disordered $530 \times 530$ nm$^2$ cross-shaped graphene conductor. The length and the nominal width of the arms amount to $L \sim 240$ nm  and $W \sim 50$ nm, respectively. A scanning electron micrograph of the actual measured sample is displayed in Fig. \ref{sample}.
The ribbon samples were fabricated from micromechanically cleaved graphene on a heavily \emph{p}-doped substrate with a 280-nm thick layer of SiO$_2$. Metallic leads to contact the graphene sheet were first patterned using standard e-beam lithography followed by a Ti(2 nm)/Au(35 nm) bilayer deposition. After lift-off in acetone, a second lithography step facilitated patterning of the GNRs.
Our measurements down to 50 mK were performed on a dry dilution refrigerator.

\begin{figure}[tbp]
\includegraphics[width=8cm]{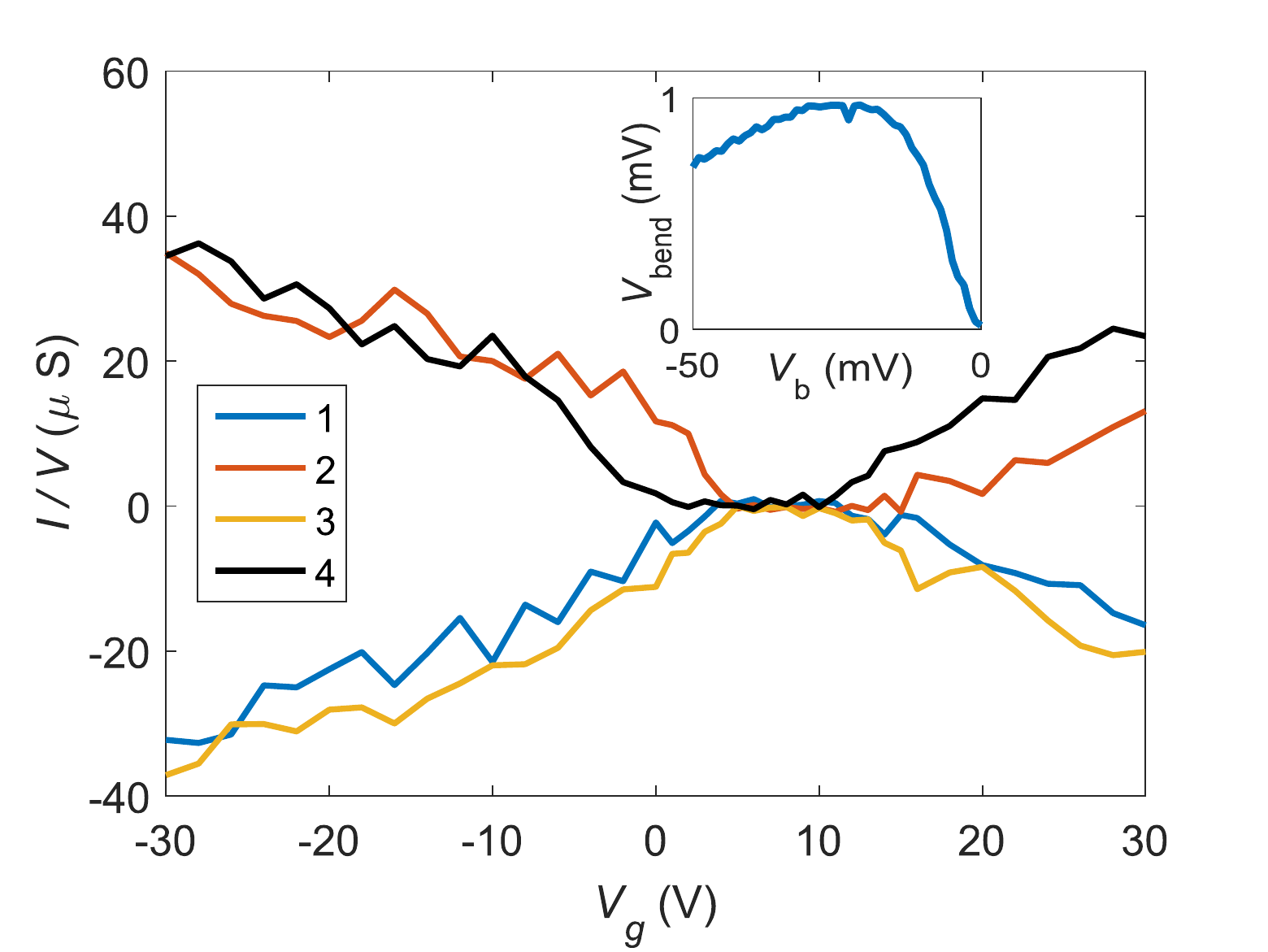}

\caption{Conductance $G=I/V$ \textit{vs.} $V_{g}$ at small $V_b$ using bias configuration C: Ingoing currents $I_2$ and $I_4$ are negative, while  $I_1 > 0$ and $I_3 > 0$. The inset at $V_g = -30$ V displays negative bend voltage $V_{21,34}$, where current is fed between terminals 4 and 3 and the voltage is measured across terminals 1 and 2.}
\label{fig:condarms}
\end{figure}

We  first characterized the sample conductances. The conductances of the arms  were derived from the data for $I/V$ in Fig.  \ref{fig:condarms} measured for the biasing configuration C, where the biasing leads 2 and 4 are seen with negative (ingoing) current, while currents in 1 and 3 are positive (see Fig. \ref{sample}). Classically, setting the potential of the center of the cross as $V/2$, we obtain the arm conductances given in Table I. The arm conductances in the properly  diffusive regime far away from the Dirac point display symmetry within approximately $\pm 6$ \% at $V_g=-30$ and $\pm 9$ \% at $V_g=-10$.  Some asymmetry in conductances is observed at $V_g \sim 0$ and $V_g \sim 15$ V.  However, the asymmetry in these regions is bias dependent and its influence becomes reduced in the constant-current-level type of correlation determinations. The conductances $G_p$ and $G_t$, defining  the behaviour in the  central region, are estimated from non-local measurements and the geometric dimensions. The size of the central region, taken as a square fitting within the middle of the cross, yields for the relative direct conductance $G_p/G_0=L/\sqrt{2}W=3.4$, where $G_0$ denotes the average arm conductance.
\begin{table} [t]
\begin{tabular}{|c | c | c | c | c |}

             \hline
              $V_g$ & Arm 1& Arm 2 & Arm 3 & Arm 4 \\
             \hline
             $-10$  V& 22 & 20 & 22 & 24 \\
              \hline
             $-30$ V & 33 & 35 & 37 & 35 \\
              \hline
 \end{tabular}
 \caption{Arm conductances at gate voltages  $V_g=-10$ V and  $V_g=-30$ V given in $\mu$S.}
\end{table}

Even though the arms of our sample are undoubtedly diffusive, the ballistic nature of the central area shows up in non-local conductance features,\textit{ i.e.} in negative bend voltage \cite{takagi1988,tarucha1992,weingart2009}. The observed bend voltage illustrated in Fig. \ref{fig:condarms}  is rather small but it indicates nevertheless that part of the charge carriers traverse the central region ballistically. The bend voltage $V_{bend}$ can be calculated using the Landauer-Buttiker theory. The result using the parametrization of Fig. 1 reads
\begin{equation}\label{bendV}
 V_{\rm{bend}}=
  \frac{G_0\,(G_t - G_p)}{ G_0\,(G_p + 3\,G_t) + 8\,G_t\,(G_p + G_t)}\,V_b.
\end{equation}
The smallness of the measured bend voltage in Fig. \ref{fig:condarms} (approximately a few per cent of bias voltage) indicates that $G_p=G_t$ within approximately $\pm 20$\%.

Our correlation measurement system operates over frequencies $f_{BW}=600-900$ MHz \cite{stick}. This frequency is typically well above any fluctuator noise due to switching in transmission eigenvalues at the contacts \cite{Laitinen2014}. Still, the
employed frequency range is low enough to correspond to zero-frequency noise because the frequency is low compared
with the internal $1/RC$ scale and the temperature.
An aluminum tunnel junction was used for calibration of the
noise spectrometers \cite{danneau2008b}. For IV curves with significant non-linearity,
$S_{nm}$ values were first derived as a function of current in the limit $I \rightarrow 0$. The resulting reading of $dS_{nm}/dI$ was converted to $dS_{nm}/dV_b$ by using the measured differential conductance $dI/dV_b$.  Note that we
always take the opposite of the cross correlations when $n \neq m$, which makes $S_{nm}$ positive as all these non-
diagonal correlations are negative in a fermionic system.

Our results on the auto and cross correlation power show linear slope with
current at small bias, which goes weakly down at large bias  where inelastic scattering starts to take place.
When inelastic processes are important (inelastic length $l_{in} \lesssim L$), shot noise in graphene is reduced by
the most strongly coupled energy relaxation processes, i.e. either by impurity-assisted acoustic phonon collisions
or by optical phonons \cite{fay2011,Betz2012,Laitinen2014,Laitinen2015}. Inelastic processes were strongest in our work near the CNP, as rather large voltages were needed for biasing.

Autocorrelation $S_{11}$ was investigated with bias $V_b$ in terminal 1 and the other terminals grounded. In this
configuration, we found $F \simeq 0.4$, close to the values reported in Ref. \onlinecite{Tan2013}
for a configuration with floating side terminals; similarly $F$ was increased near the CNP where the IV curves
are strongly non-linear. This Fano factor is higher than the universal value $F = 1/3$ for diffusive systems.  Away
from the CNP, our value of $F\simeq 0.4$ is in agreement with the theoretical results for disordered graphene
ribbons \cite{lewenkopf2008}. In Ref. \cite{Tan2013} it was concluded that these results are in accordance with
Gaussian disorder having a dimensionless strength of $K_0 \approx 10$, which meant that the conductance is
strongly affected by disorder \cite{lewenkopf2008}. However,
almost the same result is obtained in our model with diffusive arms and ballistic central region. Indeed, the
calculations for $G_p = G_t = 3.4\,G_0$ give $F_f=0.367$ for the floating side terminals and $F_g=0.394$ for the
three grounded ones \cite{suppl}.
\begin{figure}[tbp]
\includegraphics[width=8cm]{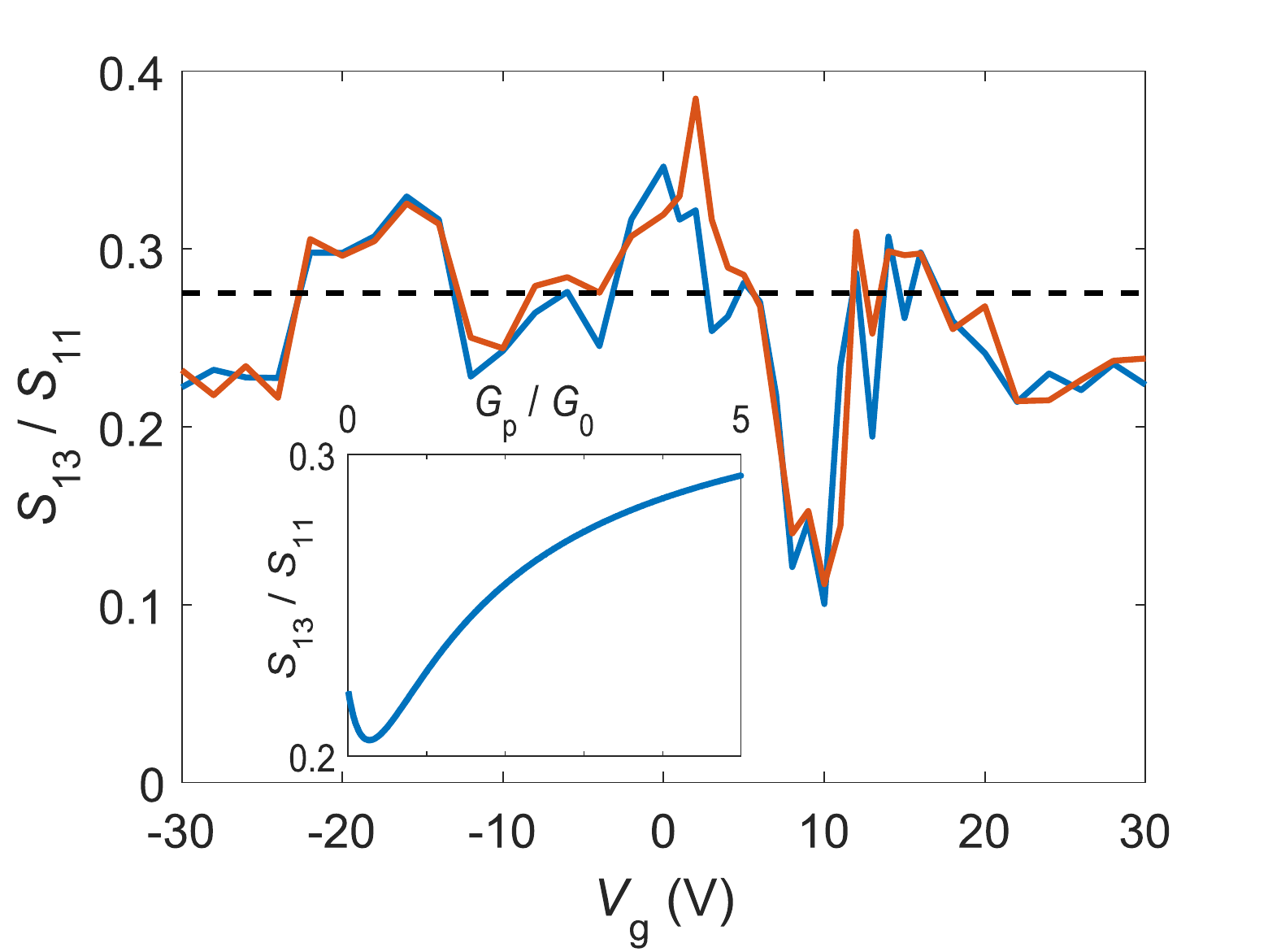}
\caption{Ratio of $S_{13}/S_{11}$  vs. $V_g$ with bias applied via terminal 1 having other terminals DC grounded. The two data sets, light and dark, relate to $V_b \lessgtr 0$, respectively: their difference is indicative of the small uncertainty in the data. The dashed line indicates the result from our model with $G_p/G_0 = 3.4$. The inset displays the calculated behavior of $S_{13}/S_{11}$ \textit{vs}. the ratio $G_p/G_0 $ ($G_p/G_t  = 1$). }
\label{cross_auto}
\end{figure}

For a symmetric diffusive cross with negligible resistance of the central region $S_{13}/S_{11}=1/3$.
Our calculated result for this ratio depends on the significance of the occupation number noise and  the inset in Fig. \ref{cross_auto} depicts the theoretical ratio as a
function of $G_p/G_0$ with $G_p = G_t$.
In the limit of $G_p \rightarrow \infty$, our theory yields the diffusive value $S_{13}/S_{11}=1/3$, while in the
limit of $G_p \rightarrow 0$ and for $G_p = G_t$, we obtain 2/9. Hence, we expect a clear deviation from the
diffusive value when the occupation-number  noise of the central region plays a role, even though no
asymmetry exists in the conduction and $G_p = G_t$ at the crossing.

A deviation from the diffusive
behavior is apparent in Fig. \ref{cross_auto}, which displays the measured ratio $S_{13}/S_{11}$ for our graphene cross.
 Our theoretical calculation for $G_p/G_0 = 3.4$ yields $S_{13}/S_{11} =
0.275$, which agrees well with the experimental data away from the Dirac point.
At $V_g=-30 \ldots -10$ V, the sample behaves as a uniform system, where current is divided nearly equally from a single
biased lead into the three other arms (see Table I). Consequently, the clear variation of $S_{13}/S_{11}$ in this regime indicates non-constant value for $G_p/G_0$ as a function of charge density, as otherwise $S_{13}/S_{11}$ should remain better as constant at $|V_g| > 20$ V.
In the range of $V_g=3  \ldots +13$ V, in particular, the
electrical transport is influenced by hopping conduction.
Near the CNP ($8 \textrm{ V} < V_g < 10$ V), we find a strong decrease in $S_{13}/S_{11}$ which cannot be accounted
for by our extended diffusive cross theory. Increased shot noise due to tunneling and localized states near the
CNP is a likely contribution to the strong decrease of  $S_{13}/S_{11}$.

\begin{figure}[tbp]
\includegraphics[width=7.7cm]{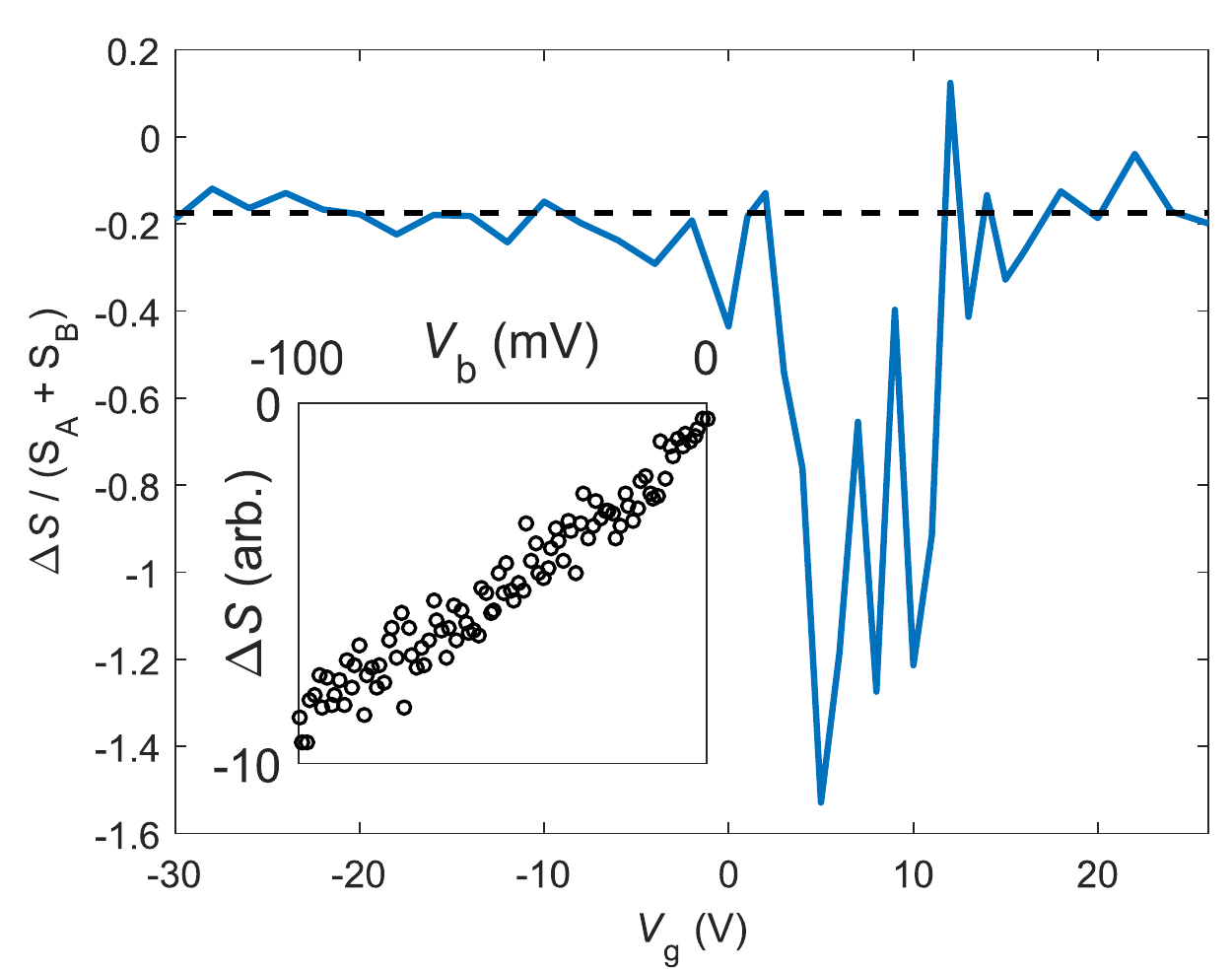}
\caption{HBT exchange correction $\Delta S$ vs. $V_g$ obtained from low-bias cross correlation experiments extrapolated to $V_{b} \rightarrow 0$. The solid line indicates our model result $\Delta S/(S_A+S_B) S = -0.175$ using $G_p/G_0 = 3.4$.
The inset displays the linear dependence of $\Delta S$ on $V_b$ measured at $V_g=-30$ V. }
\label{HBTexp}
\end{figure}

Finally, we present our data on the Hanbury -- Brown and Twiss effect.
We define the HBT exchange correction term in accordance with Ref. \cite{buttiker1997} by
$\Delta S = S_C - S_A - S_B$, where $S_A$, $S_B$, and $S_C$
denote the absolute values of the cross-correlated noise power spectra between terminals 1 and 3
in three different configurations {\it A, B} and {\it C}: in the HBT configuration $A$, ($B$), bias was applied to
terminal 2 (4) while the other terminals were connected to DC ground; in the case $C$, both 2 and 4 were biased and
1 and 3 DC-grounded.
In order to compare the experimental results more accurately with theoretical predictions , we present the scaled HBT ratio $\Delta S/(S_A +S_B)$ in Fig. \ref{HBTexp}.

\begin{figure}[tbp]
\includegraphics[width=7.7cm]{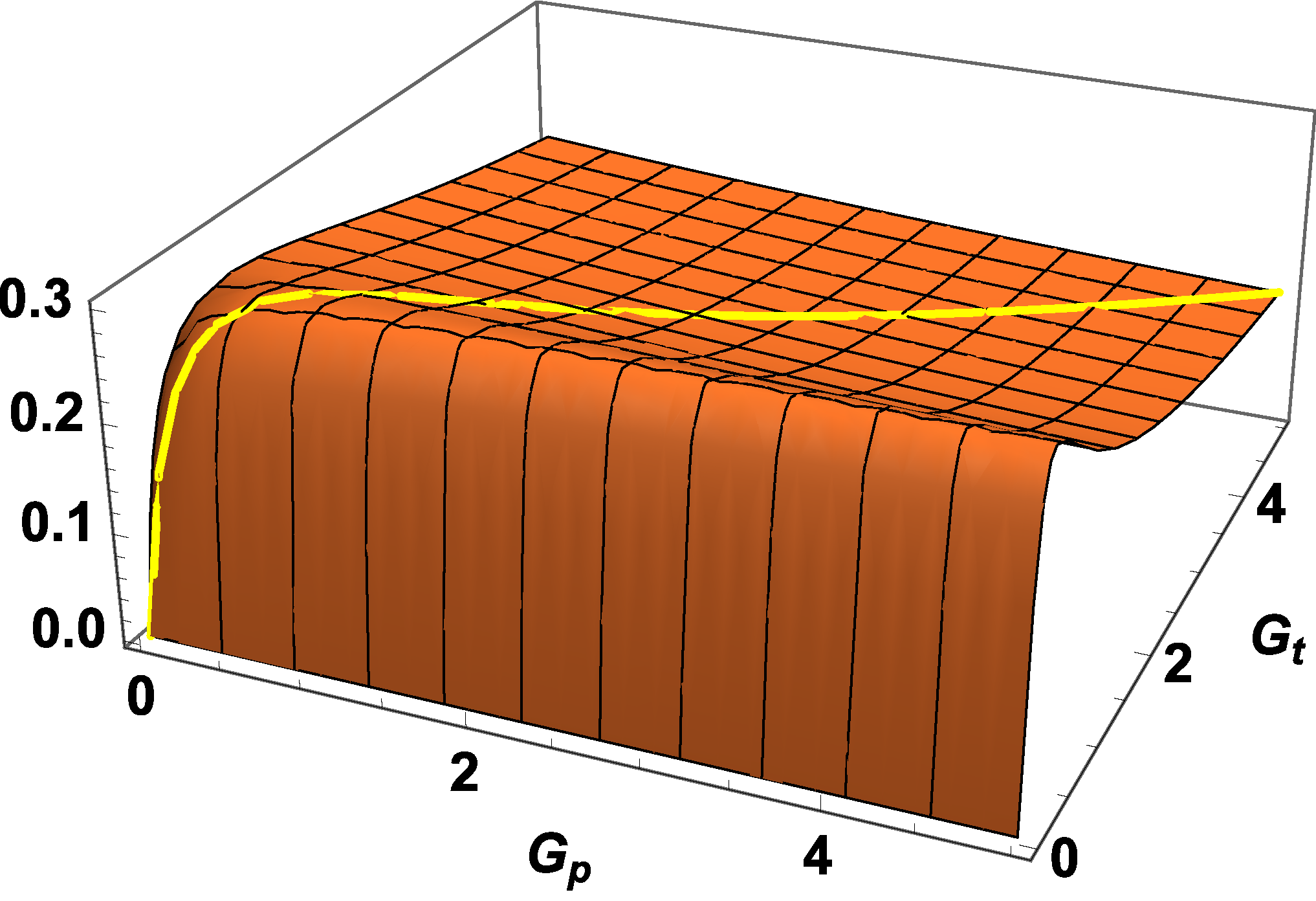}
\caption{Theoretically calculated HBT effect $\Delta S/(S_A+S_B)$ as a function of $G_p$ and $G_t$; the arm conductance is set to $G_0 = 1$. In our analysis we are using the overlaid trace for $\Delta S/(S_A+S_B)$  on the diagonal at which $G_p=G_t$. }
\label{fig:deltaStheory}
\end{figure}

  Our data display a strong modification of the HBT exchange factor near the Dirac point. The strength of this change, however, cannot be captured by our extended diffusive conductor model. In our theoretical model (see Fig. \ref{fig:deltaStheory}), $\Delta S$ is seen to vary with the ratio of $G_p/G_0$ which is likely to be modified near the CNP point. Thereby, a moderate decrease in $\Delta S$ could be understood in terms of a stronger gate dependence in the conductance in the central region compared to that in the arms.  Such a mechanism, however, would only account for values $|\Delta S| /(S_A+S_B) < 0.30$,  which falls clearly short from our measured results.

To account quantitatively for our experimental results, we have calculated the
noise in a graphene cross using the semiclassical Boltzmann-Langevin equations for
diffusive arms, which are connected via the ballistic central region. Due to the nonequilibrium distribution of
electrons in the arms and the occupation-number  noise in the central region, we obtain \cite{suppl}
\begin{multline}\label{deltaStheory}
\Delta S= -\frac{20}{3}\, G_0^2 G_t^2\,
 \frac{3\,G_0+2\,G_p+10\,G_t}{(G_0+2\,G_p+2\,G_t)^2}
 \\ \times
 \frac{G_0 G_p+2\,G_p G_t+2\,G_t^2}{(G_0+4\,G_t)^4}\,eV_b.
\end{multline}
instead of $\Delta S =0$, that is given for regular diffusive system by the semiclassical theory
\cite{Sukhorukov1999}, and $\Delta S >0$ predicted for ballistic graphene \cite{laakso2008}. The calculated result for $\Delta S/(S_A + S_B)$ is displayed in Fig. \ref{fig:deltaStheory}
on the plane spanned by $G_p$ and $G_t$. The regular diffusive behavior $\Delta S = 0$ is obtained in the limit $G_p
\rightarrow \infty$. The dashed line in Fig.
\ref{HBTexp} is obtained from Eq. \eqref{deltaStheory} using $G_p=G_t $ and $G_p/G_0 = 3.4$. The agreement between
the model and the data is excellent in the regime where the charge density in the sample  is large.

The strong growth of the absolute value of HBT exchange effect near the CNP is presumably due to Coulomb blockade
and tunneling becoming more important. For an island connected to four metallic leads by four tunnel junctions, we have measured $\Delta S/(S_A+S_B) = -1$ within 3\%. We  did also check whether enhanced electron-electron interactions
near the CNP could account for the enhanced HBT effect at small charge density. However, a calculation for the hot
electron regime (without a ballistic center) yields $\Delta S/(S_A+S_B)=-0.295$, which is clearly different from the measured results both near the CNP and far away from it.

To conclude, we have studied cross correlations in a diffusive, disordered graphene conductor where the
elastic mean free path is on the order of the feature size of the geometric layout. Even
though the non-local behavior in our graphene cross is quite weak, its presence pinpoints occupation-number  noise
that has a profound effect on the current-current correlations of the multiterminal system.
As a consequence, the noise properties of this disordered system cannot be treated by standard diffusive theories. By
inclusion of the occupation-number noise, excellent agreement is obtained between our
semiclassical model and the measured noise properties, including the Hanbury--Brown and Twiss exchange effects in
the regime where charge density is large. Near the charge neutrality point, Coulomb effects become strong and a
large HBT effect with strong fluctuations  is observed.

We thank D. Golubev, C. Flindt, G. Lesovik, F. Libisch, C. Padurariu, P. Virtanen, and T. Heikkil\"a for fruitful discussions. This work has been supported in part by the EU 7th Framework Programme (Graphene Flagship and Grant No. 228464 Microkelvin), by the Academy of Finland (projects no. 250280 LTQ CoE and 132377) and Russian Foundation
for Basic Research, Grant No. 16-02-00583-a, the
program of Russian Academy of Sciences.

\newpage
\renewcommand{\thefigure}{S\arabic{figure}}
\renewcommand{\theequation}{S\arabic{equation}}

\setcounter{figure}{0}
\setcounter{equation}{0}

\begin{widetext}

\Large{Cross correlations in disordered, four-terminal graphene-ribbon conductor: Hanbury-Brown and Twiss
exchange as a sign of non-universality of noise -- Supplementary information}

\normalsize

\section{Bend voltage and bend resistance}

We adopt the following model of the conducting cross with a ballistic central region. The four arms of the cross are assumed to be diffusive with equal conductances $G_0$, and the central region is assumed to be ballistic. More precisely, there are $2m+n$ reflectionless channels originating from each arm, and $n$ of
them go straight ahead into the opposite arm, while $m$ channels turn left and another $m$ turn right
(see Fig. 1). Hence the central region may be treated as  a four-terminal conductor with the
conductance matrix
\begin{equation}
 \hat{G} = \begin{pmatrix}
  2\,G_t + G_p & -G_t & -G_p & -G_t
  \\
  -G_t & 2\,G_t + G_p & -G_t & -G_p
  \\
  -G_p & -G_t & 2\,G_t + G_p & -G_t
  \\
  -G_t & -G_p & -G_t & 2\,G_t + G_p
 \end{pmatrix}
 \label{G}
\end{equation}
where $G_p = 2me^2/h$ and $G_t = 2ne^2/h$. If the electrical potential in arm $i$ at the crossing is
$\varphi^c_i$, the total current flowing from this arm into the other arms equals
\begin{equation}
 I_i = \sum_j G_{ij}\,\varphi^c_j
 \label{I_i-1}.
\end{equation}
On the other hand, this current is given by the Ohm's law in the diffusive arm $i$
\begin{equation}
 I_i = G_0\,(V_i - \varphi^c_i),
 \label{I_i-2}
\end{equation}
where $V_i$ is the external voltage applied to the outer end of the arm.

The two-terminal resistance between the opposite ends of the cross is calculated by setting
$V_1=V_b$, $V_2=0$, and $I_2 = I_4 = 0$. Solving Eqs. \eqref{I_i-1} and \eqref{I_i-2} for $I_1$
readily gives
\begin{equation}
 R_{2t} \equiv V_b/I_1 = \frac{G_0 + 2\,(G_p + G_t)}{G_0\,(G_p + G_t)}.
 \label{G2t}
\end{equation}

To calculate the bend voltage, it
is sufficient to substitute $V_1 = V_b$, $V_4=0$, and $I_2=I_3=0$ into the system \eqref{I_i-1} -
\eqref{I_i-2} and solve it for $V_2$ and $V_3$. As a result, one obtains
\begin{equation}
 V_{bend} = V_2 - V_3
 = \frac{G_0\,(G_t - G_p)}{ G_0\,(G_p + 3 G_t) + 8G_t\,(G_p + G_t)}\,V_b.
 \label{bend}
\end{equation}
The current flowing through arms 1 and 3 is
\begin{equation}
 I_1 = -I_3 = \frac{4G_0\, G_t\,(G_p + G_t)}
                   { G_0\,(G_p + 3 G_t) + 8G_t\,(G_p + G_t)}\,V_b,
 \label{I1}
\end{equation}
and the bend resistance equals
\begin{equation}
 R_b \equiv (V_2 - V_3)/I_1
 = \frac{1}{4}\,\frac{G_t - G_p}{G_t\,(G_p + G_t)}.
 \label{Rb}
\end{equation}

\section{Noise and exchange effect}

The noise in the systems results from two sources. One of them is impurity scattering of nonequilibrium
electrons in the diffusive arms, and the other one is the noise in the central ballistic region. Though
ballistic conductors with equilibrium electrodes do not generate noise under an applied voltage at zero
temperature, they produce it if the electron distributions in the electrodes are nonequilibrium. Here,
the diffusive arms serve as nonequilibrium electrodes for the central four-terminal ballistic conductor,
and therefore its contribution to the noise must be also taken into account. The Langevin equations
for the fluctuations of current in the diffusive arms are of the form
\begin{equation}
 \delta I_i = \delta I_i^{ext}(x_i) + LG_0\,\frac{d\delta\varphi_i}{dx},
 \label{dI-1}
\end{equation}
where $\delta I_i^{ext}$ is the extraneous Langevin current, $x$ is the coordinate along the arm,
$L$ is its length, and $\delta\varphi_i(x)$ is the fluctuation of electric potential in this arm.The
correlation function of the Langevin currents is
\begin{equation}
 \langle\delta I_i^{ext}(x)\,\delta I^{ext}_j(x')\rangle
 = 4\,\delta_{ij}\,\delta(x-x')\,LG_0 \int d\varepsilon\,f_i(x,E)\,[1-f_i(x,E)].
 \label{<dIdI>}
\end{equation}
An integration of Eq. \eqref{dI-1} over $x$ with the condition $\delta\varphi_i(0)=0$ brings it to the form
\begin{equation}
 \delta I_1 = G_0\,\delta\varphi^c_i + \int_0^L \frac{dx}{L}\, \delta I_i^{ext}(x),
 \label{dI-2}
\end{equation}
where $\delta\varphi^c_i \equiv \delta\varphi_i(L)$ is the potential fluctuation at the crossing. But on the other hand, the fluctuation of current flowing from arm $i$ into the rest of arms equals
\begin{equation}
 \delta I_i = \sum_j G_{ij}\, \delta\varphi^c_j + \delta\tilde I_i^{ext},
 \label{dI-3}
\end{equation}
where $\delta\tilde I_i^{ext}$ are extraneous random currents generated at the crossing due to the nonequilibrium distribution of incident electrons with the correlation function
\begin{equation}
 \langle\delta\tilde I_i^{ext}\,\delta\tilde I_j^{ext}\rangle
 =
 2\,G_{ij} \int dE\,
 [ f^c_i\,(1 - f^c_i) + f^c_j\,(1 -f^c_j)],
 \label{<dIdI-2>}
\end{equation}
where $f^c_i(E)$ is the distribution function of electrons in arm $i$ at the crossing. Its
values may be obtained from a system of equation similar to Eqs. \eqref{I_i-1} and
\eqref{I_i-2} with $f^c_i$ in place of $\varphi^c_i$ and $f_0(E-eV_i)$ in place of $V_i$,
where $f_0(E)$ is the Fermi distribution function. The distribution function of electrons
in the arms is governed by simple diffusion at a given energy and presents a linear
combination of distributions at its ends
\begin{equation}
 f_i(x,E) = \left( 1 - \frac{x}{L}\right) f_0(E-eV_i) + \frac{x}{L}\,f_i^c(E).
 \label{f(x)}
\end{equation}
The system of equations \eqref{dI-2} and \eqref{dI-3} has to be solved for $\delta I_1$ and $\delta I_3$,
and then the correlation function $S_{13} = -\langle\delta I_1\,\delta I_3\rangle$ has to be calculated using
Eqs. \eqref{<dIdI>} and \eqref{<dIdI-2>}.

First of all, we calculate the two-terminal Fano factor for the case where the current flows only through
arms 1 and 3, whereas side arms 2 and 4 are floating. Hence one has to set $V_1=V_b$,
$V_3=0$ and $I_2 = I_4 =0$ for the averages and $\delta V_1 = \delta V_3 = \delta I_2 = \delta I_4 =0$ for the
fluctuations. This gives us the Fano factor in the form
\begin{equation}
 F_f \equiv S_{11}/|eI_1| =
 \frac{2}{3}\,(G_p + G_t)\,
 \frac{6\,G_0^2 + 9\,G_0\,(G_p + G_t) + 4\,(G_p + G_t)^2}
      {[G_0 + 2\,(G_p + G_t)]^3}.
 \label{Fano-fl}
\end{equation}
\begin{figure}
\scalebox{0.7}{\includegraphics{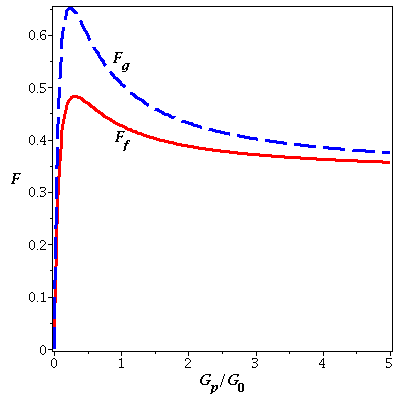}}
\caption{Dependence of the Fano factor on the ratio between the conductances of the crossing point
 and the arms for $G_p=G_t$. The solid line corresponds to $F_f$ for the floating side terminals, and the dashed line corresponds to $F_g$ for the three grounded terminals.}
\label{Fano_fig}
\end{figure}
It is easily seen that $F_f$ depends only on the ratio between $G_p + G_t$ and $G_0$. It tends to zero as for a
purely ballistic system when this ratio is small and approaches the 1/3 value for a diffusive conductor when
it is large. As shown in Fig. 
S1, it passes through a maximum $F_f \approx 0.48$ at
$(G_p + G_t)/G_0 = (\sqrt{5}-1)/2 \approx 0.62$ and equals 0.367 for $G_p = G_t = 3.4\,G_0$. This is larger
than the shot noise for diffusive contacts with purely elastic scattering, but somewhat smaller than the noise in the hot-electron regime.

In a configuration where the voltage is applied to terminal 1 and the rest of terminals are grounded, one
sets $\delta V_i=0$ for all $i$. The general expression for the Fano factor is too cumbersome, and we
present it here only for the particular case of $G_p=G_t$, where it reads
\begin{equation}
 F_g  \equiv S_{11}/|eI_1| =
 \frac{2}{3}\,G_p
 \frac{18\,G_0^2 + 45\,G_0 G_p + 32\,G_p^2}
      {(G_0 + 4\,G_p)^3}.
 \label{Fano-gr}
\end{equation}
The $F_g(G_p/G_0)$ curve is similar in shape to $F_f(G_p/G_0)$ but lies higher and reaches its maximum
$F_g=0.65$ at $G_p/G_0 \approx 0.24$. For the particular values $G_p = G_t = 3.4\,G_0$, $F_g$ equals 0.394.

In a similar way, one calculates $S_{13}^A$ for $V_1=V_3=V_4=0$ and $V_2=V_b$, $S_{13}^B$ for $V_1=V_2=V_3=0$
and $V_4=V_b$, and $S_{13}^C$ for $V_1=V_3=0$ and $V_2=V_4=V_b$. The resulting exchange term in the noise equals
%
\begin{equation}
 \Delta S \equiv S_{13}^C - S_{13}^A - S_{13}^B \\
 = -\frac{20}{3}\,
 \frac{G_0^2\,G_t^2\,(10\,G_t + 2\,G_p + 3\,G_0)(G_0\,G_p + 2\,G_p\,G_t + 2\,G_t^2)}
      { (G_0 + 2\,G_p + 2\,G_t)^2(G+0 + 4\,G_t)^4 }\,eV_b.
 \label{DS}
 \end{equation}

\end{widetext} 

\end{document}